\documentclass[11pt,a4paper]{article}

\renewcommand{\author}{Emil Khalisi}
\newcommand{\titel}{Joshua's Total Solar Eclipse at Gibeon}
\newcommand{\version}{Version 1.83}
\renewcommand{\date}{\today}

\usepackage{hyperref}
\hypersetup{pdfauthor={Emil Khalisi}}
\hypersetup{pdftitle={\titel }}
\hypersetup{pdfsubject={Online publication, arXiv, \version\ of \date\ }}
\hypersetup{pdfkeywords={Solar eclipse, Earth's rotation, astronomical dating,
   Gibeon, Palestine, Book of Joshua}}
\hypersetup{colorlinks=true,citecolor=blue}

\usepackage[T1]{fontenc}        
\usepackage{mathptmx}           
\usepackage{pifont}             
\usepackage[scaled=0.92]{helvet} 

\usepackage[british]{babel}     
\usepackage{amsmath}            
\usepackage{microtype}          

\usepackage{multicol}           
\usepackage{paralist}           

\usepackage{calc}
\usepackage[a4paper]{geometry}  
\usepackage{scrextend}           
\changefontsizes{10pt}

\usepackage{titlesec}
\titleformat*{\section}{\large\bfseries}
\titleformat*{\subsection}{\normalsize\bfseries}

\usepackage{graphicx}           
\usepackage{float}              
\usepackage{caption}            
\usepackage{fancyhdr}
\renewcommand{\headrulewidth}{0.4pt}
\usepackage{colortbl}             
\definecolor{grey20}{RGB}{208,208,208}

\usepackage{url}
\usepackage[numbers]{natbib}      
\begin{document}



\fancyhead{}
\fancyhead[LO]{%
   \footnotesize \textsc{In original form published in:}\\
   {\footnotesize Habilitation at the University of Heidelberg }
}
\fancyhead[RO]{
   \footnotesize {\tt arXiv:****.***** [physics.hist-ph]}\\
   \footnotesize {Date: 18th February 2021}%
}
\fancyfoot[C]{\thepage}

\renewcommand{\abstractname}{}

\twocolumn[
\begin{@twocolumnfalse}

\section*{\centerline{\LARGE \titel }}

\begin{center}
{\author \\}
\textit{D--69126 Heidelberg, Germany}\\
\textit{e-mail:} \texttt{ekhalisi[at]khalisi[dot]com}\\
%
\end{center}


\vspace{-\baselineskip}
\begin{abstract}
\changefontsizes{10pt}
\noindent
\textbf{Abstract.}
We reanalyse the solar eclipse linked to the Biblical passage about
the military leader Joshua who ordered the sun to halt in the midst
of the day (Joshua 10:12).
Although there is agreement that the basic story is rooted in a real
event, the date is subject to different opinions.
We review the historical emergence of the text and confirm that the
total eclipse of the sun of 30 September 1131 BCE is the most likely
candidate.
The Besselian Elements for this eclipse are re-computed.
The error for the deceleration parameter of Earth's rotation,
$\Delta T$, is improved by a factor of 2.

\vspace{\baselineskip}
\noindent
\textbf{Keywords:}
Solar eclipse,
Earth's rotation,
Gibeon,
Palestine,
Book of Joshua.
\end{abstract}

\centerline{\rule{0.8\textwidth}{0.4pt}}
\vspace{2\baselineskip}

\end{@twocolumnfalse}
]




\section{Introduction}

Eclipses provide magnificent natural spectacles, but only the type
of a \emph{total} solar eclipse produces darkness as deep as in the
night, almost instantly, with stars appearing.
Other types of eclipses (annular or partial) may be great events,
but they would not compete with those total ones that leave behind
a breathtaking once-in-a-lifetime experience to the observer.
Such an eclipse occurs rarely at a given place, about three times
in a millennium on average, though the time interval between two
subsequent occurrences varies a lot in specific cases.

The cause of an eclipse remained unknown for most people in ancient
times, and so it was for the composers of the Old Testament.
In particular, spiritual persons took it as a sign of supernatural
power or messages from God.
Various prophets, most of whom never had the vaguest notion of
astronomical cycles, intensified the superstitious fear of divine
dissatisfaction with the human practices.
The total darkness at an unexpected time of day was terrifying
so much that the affected changed their behaviour.

The best known example of a positive aftermath was the so-called
``Thales' eclipse'', in 585 BCE, when Medes and Lydians hurried to
make peace in Asia Minor after having combated for years.
More than a half millennium before there was another battle near
Gibeon in Palestine fought between the Jewish army and an alliance
of five Amorite kings.
This account is handed down in the scriptures of the Old Testament,
but heavily steeped in religious propaganda.

In this paper, we will present the example how a total solar eclipse
gave rise to taking advantage of the enemy's confusion during that
moment.
We will sum up the roots of the Jewish history and then unfurl the
relevant passage about the eclipse.
Additionally, we use it to confine the deceleration parameter of the
Earth's rotation, $\Delta T$, more precise than previously.

\section{Historical Evidence for the Early Jews}

The whole Book of Joshua comprises 24 chapters, and it appears as
a work of many anonymous authors, but attributed to Joshua himself.
Almost all scholars agree that the first 11 chapters were written
in late 7th century BCE.
They were not completed until after the capture by the Neo-Babylonian
Empire in 586 BCE, and incorporated into the Bible in a revised
version possibly after the return from the Babylonian exile in 538
BCE \cite{creach:2003}.
The story was intended to readers of the 7th and 6th century BCE
when old traditions from the time of Exodus were revived.
Furthermore, the belief in god, social stratification as well as
ethnic groups were political issues at that time \cite{nelson:1997}.

\begin{figure}[t]
\centering
\includegraphics[width=\linewidth]{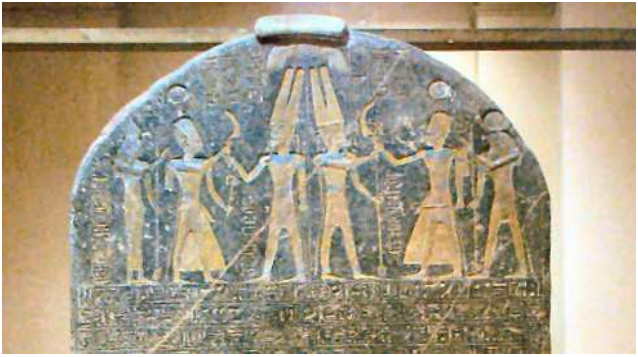}
\caption{Upper part of the Merneptah Stele in the Egyptian Museum
    in Cairo \cite{w-en}.}
\label{fig:merneptah}
\end{figure}

From the archaeological point of view, the Jewish culture starts in
the 12th century BCE.
The oldest known reference to the word ``Israel'' is given on the
victory stele of the Egyptian pharaoh Merneptah discovered at
Thebes, now in Cairo (Fig.\ \ref{fig:merneptah}).
It is the sole non-Biblical note until the 9th century BCE belonging
to an Assyrian inscription \cite{gertz:2019}.
The reign of Merneptah lasted from $\approx$1213 to 1203 BCE
depending on uncertainties inherited from previous chronological
difficulties.
Merneptah was the son and successor of Ramses II, who played an
important role in the old Egyptian history.
At that time, the term Israel was not understood as a geographical
place but rather a group of people.
It could also refer to the name of the leader of a nomadic group
working in Egypt.

The roots of Jewish thoughts may go back one or two centuries
earlier in time.
The concept of a monotheistic religion with the sun as the only
deity was introduced by Akhenaten in the 14th century BCE
\cite{khalisi-egypt}.
After the death of this pharaoh, the Egyptian priesthood returned
to their former customs of many gods, but a small group might have
maintained one or the other idea.
It is not unusual in history that new mindsets take some
generations to thrive and prosper.
A similar evolution happened, for instance, to Christianity:
after the execution of Jesus it took more than 100 years till
Eastern was celebrated.
This ``celebration'' was nothing more than just a commemoration day
for a group split off from the Jews calling themselves Christians.
It was only in 313 CE, when the Edict of Milan by Constantine the
Great (ca.\ 272--337) appointed all religious minorities equal,
and the small community of Christians benefited most from this
Edict.
Also, Buddhism began to disseminate about 100 to 150 years after
the death of its teacher (5th century BCE),
and Confucianism was elevated to state religion in China during
the Han dynasty, four centuries after the lifetime of the
philosopher Confucius (551--479 BCE).
Islam too expanded in the first decades geographically, but it
achieved its cultural prosperity during the second dynasty, the
Abbasids, from the mid-8th century onwards.
Many prophet-based religions gain their momentum a considerable
time after the flourishing of their founder.
Especially, a ``miracle'' would boost legends about him:
the more mysterious, the greater the effect.
Judaism makes no exception.

According to the Old Testament, the ``Israelites'' became a unity
during their stay in Egypt.
Historical research rates all early descriptions about that very
critical \cite{sparks}.
Scholars agree that history takes notice of them at the end of the
Bronze Age after the collapse of the Hittite and the Egyptian
Empires.
Sea peoples were ravaging the Eastern Mediterranean then,
destabilising the whole region which is now called Palestine.
It was a period of tremendous violence in the Eastern Mediterranean.
Thereafter, the Israelites still remained a local minority.
Evidence for the later Biblical Empires of the Kings David and
Solomon at about 1000 BCE could never be found \cite{gertz:2019}.
These could have been local princes, at best, who attained a
certain degree of self-government inside that region.
The first king of Northern Israel is said to be Jeroboam
(reign ca.\ 926--907), but the historicity relies on interpretations
of one or two indications rather than on archaeological evidence.

In all, we just do not know when the Jewish religion came into
being and formed an ideological community.
Moses, who is said to have led a large group of people out of Egypt,
is not datable.
He is not affirmed as a historical person at all, since he is
nowhere mentioned in contemporary records outside those holy
scriptures.
Furthermore, he would not be the real author of the starting
chapters of the Bible, anyway.
All statements about his life are annexations of much later times.
Many small elements of the whole Jewish story show motives from
legends of diverse Mesopotamian cultures and seem to be ``borrowed''.

%
\fancyhead{}
\fancyhead[C, C]{\footnotesize \itshape E.\ Khalisi (2021): \titel}
\renewcommand{\headrulewidth}{0pt}

\section{The Account on the Eclipse}

The five Books of Moses are succeeded by the Book of Joshua.
The author is unknown, too.
While Moses is said to have died before the arrival in Canaan, it
was Joshua who conquered it.
Joshua, the son of Nun, is depicted as Moses' servant who was
appointed his successor and military leader later on.
The entire book is very likely a compilation from various eras,
because Joshua could not have taken part in all the combats
attributed to him \cite{sawyer:1972}.
Most important here is the Battle of Gibeon that contains a clear
report of a solar eclipse.

The narrative stands in the context of the conquest of the land
after the Exodus from Egypt.
The backstory is that the ancient city of Gibeon, about 10 km to
the northwest of Jerusalem made peace with the foreign settlers who
arrived some time before.
Thereupon five Amorite kings under the lead of the Canaanite king
of Jerusalem laid siege to Gibeon.
The defenders asked their ally Joshua for help.
He ordered a nocturnal march through rough terrain and led his
army to the gates of that city.
The assault began in the early morning.
Joshua succeeded in overcoming the five kings.
The defeated fled to the nearby settlement of Beth-Horon, and the
Jews chased them.

The eclipse miracle is said to have occurred at noon time.
According to the account in Chapter 10:11--14, Joshua's army
benefited from the confusion to commit a massacre among the enemies
\cite{bibleserver}:
\begin{quote}
(11) As they fled before Israel on the road down from Beth Horon to
Azekah, the Lord hurled large hailstones down on them, and more of
them died from the hail than were killed by the swords of the
Israelites.
(12) On the day the Lord gave the Amorites over to Israel, Joshua
said to the Lord in the presence of Israel:
``Sun, stand still over Gibeon, and you, moon, over the Valley of
Aijalon.''
(13) So the sun stood still, and the moon stopped, till the nation
avenged itself on its enemies, as it is written in the Book of
Jashar.
The sun stopped in the middle of the sky and delayed going down
about a full day.
(14) There has never been a day like it before or since, a day when
the Lord listened to a human being.
Surely the Lord was fighting for Israel!
\end{quote}

The technical term ``eclipse'' is never used.
The events are rather described from the viewpoint of a believer
in miracles.
The writer seems not to have experienced a total solar eclipse
himself.
He writes about the happenings from hearsay.
It seems obvious that he would hardly know how to put the strange
phenomenon in a correct expression, especially, when not being
aware of the mechanism behind it.
Instead, priority is given to the theological message:
faith in God would be the reason for the victory.

\section{Interpretations}

The interpretations of the standing still are variegated:
a literal understanding, a meteorite, a cult related to the sun and
moon, a description of ``opposition'' in astrological terms, or a
storm blotting out the sun and moon \cite{day:2007}.
It has also been taken as a ``prayer'' that the sun not dissipate
the morning mist so that surprise could be preserved
\cite{nelson:1997}.
All these views are, simply said, not tenable.

The first to interpret the expression ``the sun stood still'' as an
eclipse was the linguist Robert Wilson (1856--1930)
\cite{wilson:1918}.
He grasped the strange line that the sun ``delayed going down about
a full day'' that the day decayed in two parts:
after the usual sunrise in the morning the loss of light happened
at midday.
When the light returned after the eclipse, a second day began while
the sun did not move forward.
The weird day was ``cut'' in two parts.

When imagining the scene today, the unprepared observer will be
petrified, and his full attention will be fixed to the black sun in
the sky.
Those few minutes of totality leave behind such a hypnotising effect
that the duration of events is overrated.
In a state of trepidation minutes appear like hours.
The subsequent transmission will stretch the story and modify the
comprehension.
These fallacious perceptions of an extreme duration have been
documented more than a dozen times in history, even in our modern
era, cf.\ the eclipse of 1860 at Dongola (Dunqula) in Sudan, when
locals alleged the darkness would have lasted for 2 hours
\cite{mahmoudbey}.
Passing on such information, an uninvolved understands the
description as a separation between two different days:
the sun was still standing in the sky where ``it went down'' and
re-appeared from that same spot \cite{sawyer:1972}.
Thus, the passage in Joshua 10 looks very much like one of these
numerous examples and should be taken as an evidence for totality.

Before turning to the selection of suitable eclipses for the
battlefield, it may be noteworthy that the writer of the text in
Joshua 10 does not relate the role of the moon directly to the sun.
Both luminaries are treated as separate objects.
There is no indication that he understood their interaction when
the phenomenon occurred.
This arouses the suspicion that the moon was added at another time
to magnify the importance of Joshua's miracle.
In our extensive study on eclipses we never found any clues from
any culture that someone prior to the 7th century BCE comprehended
that the moon was responsible for a solar eclipse
\cite{finsternisbuch}.


\begin{table}[t]
\caption{List of all eclipses between 1300 and 1000 with mag $>$0.8
    for Beth Horon (LT = UT + 2$^{\rm h}$ 7$^{\rm m}$).
    Types: A = annular, T = total, H = hybrid.}
\label{tab:beithoron}
%
\centering
\begin{tabular}{lcccr}
\hline
\rowcolor{grey20}
  Date [BCE]  &  LT   & Type& Magn. & $\odot$-height \\ 
\hline
 1281 Apr 14  & 07:18 &  A  & 0.803 & 21.5$^{\circ}$  \\
 1261 Sep 27  & 15:58 &  A  & 0.870 & 23.6$^{\circ}$  \\ 
 1258 Jul 27  & 10:42 &  T  & 0.832 & 72.5$^{\circ}$  \\
 1247 Dec 30  & 11:36 &  A  & 0.933 & 34.7$^{\circ}$  \\
 1223 Mar 05  & 13:36 &  T  & 0.850 & 42.3$^{\circ}$  \\
 1207 Oct 30  & 16:35 &  A  & 0.950 &  7.2$^{\circ}$  \\ 
 1197 Oct 09  & 08:08 &  A  & 0.870 & 29.2$^{\circ}$  \\
 1192 Jan 21  & 13:19 &  A  & 0.884 & 33.0$^{\circ}$  \\
 1183 Jan 12  & 08:47 &  A  & 0.959 & 18.3$^{\circ}$  \\ 
 1157 Aug 19  & 08:35 &  T  & 0.986 & 43.1$^{\circ}$  \\ 
 1132 Apr 17  & 14:50 &  A  & 0.898 & 40.4$^{\circ}$  \\ 
 1131 Sep 30  & 12:30 &  T  & 1.007 & 57.4$^{\circ}$  \\ 
 1129 Feb 14  & 09:54 &  H  & 0.949 & 32.1$^{\circ}$  \\
 1124 May 18  & 10:07 &  T  & 0.864 & 64.4$^{\circ}$  \\
 1103 Sep 21  & 08:53 &  H  & 0.963 & 41.7$^{\circ}$  \\
 1091 Aug 09  & 16:51 &  A  & 0.977 & 22.1$^{\circ}$  \\ 
 1090 Dec 25  & 10:43 &  A  & 0.883 & 32.2$^{\circ}$  \\
 1084 Mar 27  & 11:18 &  T  & 0.870 & 55.5$^{\circ}$  \\
 1078 May 20  & 09:00 &  A  & 0.908 & 51.5$^{\circ}$  \\
 1068 Oct 23  & 14:27 &  T  & 0.932 & 32.9$^{\circ}$  \\
 1063 Jul 31  & 05:48 &  T  & 0.877 & 10.9$^{\circ}$  \\ 
 1062 Jan 25  & 07:02 &  A  & 0.812 &  0.6$^{\circ}$  \\ 
 1060 May 30  & 18:22 &  A  & 0.808 &  1.5$^{\circ}$  \\ 
 1041 Nov 23  & 07:26 &  T  & 0.982 & 11.4$^{\circ}$  \\ 
 1035 Jan 26  & 10:14 &  A  & 0.839 & 31.0$^{\circ}$  \\
 1012 May 09  & 18:00 &  T  & 0.902 &  3.3$^{\circ}$  \\ 
\end{tabular}
\end{table}


\begin{figure*}[t]
\centering
\includegraphics[width=\linewidth]{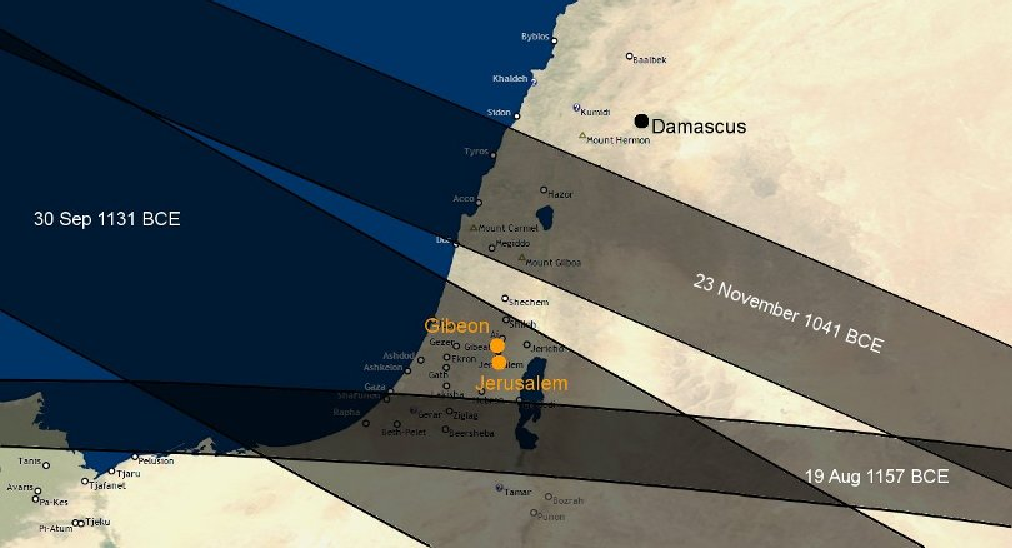}
\caption{Three suitable eclipses for the Battle of Gibeon.}
\label{fig:gibeon}
\end{figure*}


\section{Possible Eclipse Dates}

The total phase of the eclipse is preceded by a partial phase,
but the sun's disc is of such great luminousness that its
unobscured portion does not leave a noticeable effect.
A casual observer would discern changes in the illumination only
when the sun is covered more than 75\% or so,
i.e.\ in the final $\approx$15 minutes before totality.
For an impenetrable darkness a full coverage would be necessary.

Table \ref{tab:beithoron} lists all eclipses with magnitude
larger than 0.8 that principally could have been visible in
Palestine between 1300 and 1000 BCE.
Not all of them must have been really noticed because of weather
conditions.
It is a gross error to assume that a particular eclipse was seen
just because its path is sketched on a map.
Thick clouds would not change the ambient light level much from an
eclipse of mag = 0.99.
Thus, from statistical considerations one third of the candidates
could be abandoned, in principle, not knowing which.
It is more instructive to look for specific circumstances like
season of the year, height of the sun in the sky, or features of
the landscape.

In search for a feasible solar eclipse for Joshua's ``miracle''
we single out only three being intriguing:
19 August 1157, 30 September 1131, and 23 November 1041 BCE
(Fig.\ \ref{fig:gibeon}).
The first eclipse rolled by 50 kilometres south of Gibeon and had
its maximum at 8:35 a.m.\ local time.
On the supposed battlefield it would have reached a magnitude of
0.986.
Because of the track running along the line of latitude in west-east
direction, it could never reach as far north as 31.5$^{\circ}$ N
whatever displacement of $\Delta T$ is used.

The second event was more impressive:
total in Gibeon with a duration of 2 min 47 sec for the implemented
$\Delta T$ in Espenak's eclipse catalogue \cite{espenak}.
Shifting the totality zone just a few seconds to the east, it could
provide a maximum duration of 4 minutes at the central line.
Moreover, this eclipse occurred when the sun was at its summit at
noon time.
The culmination would give a broader meaning for the phrase
``the sun stood still'', but it is very doubtful that the author of
the ancient text could have considered an astronomical
understanding.

The third notable event happened on 23 November 1041 BCE, however,
the time of day was immediately after sunrise.
It does not support the logic, see arguments by John Sawyer below.

Two recent publications claim the annular eclipse of 30 October
1207 BCE be the one in question \cite{humphreys, avner:2019}.
We disregard the arguments proposed there, because a good number of
items concerning the historic facts turn out inconclusive.
Both papers get their most important pillar from the Merneptah Stele
as a ``proof'' for the Israelites having already entered Canaan.
Other issues get lost in trifling matters comparing linguistical
parallels from ``Biblical poetry''.
The perception of eclipses or the aftermath of the battle are not
addressed within the historical context.

\section{Extrapolated Delta-T}

The only eclipse that merits consideration is that of 30 September
1131 BCE, as John Sawyer and Richard Stephenson realised earlier
\cite{sawyer:1972, stephenson:1975}.
The exact location of the course can be adjusted to the width of
the totality zone.
Its hard margins confine the tolerance for determining the clock
error $\Delta T$.

We base our calculation of the Besselian Elements on the ephemeris
JPL DE406 for the time of greatest eclipse (TT = 18:03:30.7).
This starting point is the same as used by Fred Espenak
\cite{espenak}.
Our mathematical formulae to obtain the results will be published
in a supplement to the main work soon \cite{finsternisbuch}.
The computation produces slightly different values (Table
\ref{tab:belements}).
The deviations may be owed to constants for the sun, moon, and
earth (radii, distance data, flattening factor of earth's body,
atmospheric refraction).
Furthermore, Espenak seems to prefer the polynomial form of the
Elements while ours originate from direct computation of the
ephemeris.
However, the disparity affects the outcome less than 1 minute of
time and can be neglected.

Without \textit{a priori} fixing the acceleration parameter for the
earth's rotation, we find that totality passed by the site of
Beth Horon when
$\Delta T$ lies in the range of 27,644 s $< \Delta T <$ 28,041 s.
Espenak provides a mean $\Delta T =$ 27,663 s $\pm$897 s for this
eclipse.
Our values narrow down the uncertainty by a factor of 2.

The important aspect is that the average $\Delta T$ is still in
accord with the extrapolation of the parabolic formula given by
Morrison \& Stephenson \cite{morrison-stephenson:2001}:\\
\centerline{$\Delta T = -20 + 32 \cdot t^2 \;$ ,}
with $t$ being the number of centuries before 1800, i.e.\
$t =$ (1820 $-$ year)/100.
As explained in earlier papers published on {\tt arXiv}
(e.g.\ \cite{khalisi-egypt}), the extrapolation gives excellent
results for many other eclipses in the past.
Of course, this is no proof for those historical eclipses having
happened at exactly that location in question, but a strong argument
in their favour.
The formula gives a mean long-term trend from the latest fixpoints
found from Babylonian and Chinese eclipse timings.
So far, the oldest secured record on totality is of 17 July 709 BCE.
Accepting the eclipse of Gibeon as secured, too, it would shift the
earliest fixpoint by more than four centuries back in time.

\begin{table}[t]
\caption{Comparison of Besselian Elements for the eclipse of 1131 BCE
   at Terrestrial Time (TT) of the maximum.}
\label{tab:belements}
\vspace{-1ex}
\centering
\begin{tabular}{lrr}
\hline
\rowcolor{grey20}
Bess.\ El.\ &\multicolumn{1}{c}{\cellcolor{grey20}this work}
                &\multicolumn{1}{c}{\cellcolor{grey20}\cite{espenak}} \\
\hline
TT $=$      &\multicolumn{2}{c}{18:03:31} \\
$x =$       &  0,163 79 & 0,164 04 \\
$y =$       &  0,500 20 & 0,500 40 \\
$d =$       &  1,337 29 & 1,337 33 \\
$\mu =$     & 92,302 16 & 92,301 65 \\
$f_1 =$     &  0,270 85 & 0,270 84 \\
$f_2 =$     &  0,269 51 & 0,269 49 \\
$l_1 =$     &  0,536 88 & 0,537 15 \\
$l_2 =$     & -0,009 15 & -0,008 96 \\
\end{tabular}
\vspace{-1ex}
\end{table}

Unfortunately, there is no independent source about totality for
the eclipse at Gibeon.
Since the scriptures were compiled hundreds of years later, there
is the possibility that the information was taken from another
report elsewhere and added to emphasise the importance of the
battle.
This is the weak point of the account:
the writer was neither witness of the events nor contemporary.
In fact, the eclipse could have been seen at the battlefield and
it did change the balance of powers, but a very high magnitude of
obscuration would do the same job, too.
Totality itself cannot fully be verified as long as the Biblical
record remains the sole kind of ``evidence''.

\section{Discussion}

There are suggestions that the eclipse itself has nothing to do with
the battle at Gibeon.
However, John Sawyer gives three arguments for its historical
feasibility \cite{sawyer:1972}:
\begin{compactenum}
\item The time at noon fits well into the account.
The attack began in the morning, and, after several hours of
combating, the eclipse occurred redounding the powers to the
advantage of the Israelites.
\item Military campaigns were avoided in the months of mid-summer
(as the first possible eclipse of 1157 BCE was) for fear of drought.
The event of 1131 BCE points to an assault in autumn which seems
more reasonable from the strategic point of view.
\item Archaeological evidence shows that substantial remains at
Gibeon belong to the Iron Age of the late 12th century, when it
became a ``large, fortified city, governed by elders, and allied
with three other cities'' \cite{pritchard}.
It was a thorn in the side of its neighbours that such a wealthy
city like Gibeon made peace with the Jews, so the five kings laid
siege to it.
The kings would not campaign for it, if the site was of lesser
value.
\end{compactenum}

The third eclipse of 1041 BCE occurred right after sunrise in late
November.
The time of day does not agree with a battle, as the warriors
would not even have begun the combat.
All other events in Table \ref{tab:beithoron} we rate poor.

After the eclipse of 1131 BCE there was no totality to be observed
in Jerusalem and its vicinity for the next 700 years.
That long interval may foster the explanation why this particular
event left a lasting memory for the Jews and came to be re-iterated
very often till it entered the status of a ``miracle far back in
time''.
For example, the eclipse mentioned in Amos 8:9 \emph{could} refer
to that same incident, though the eclipse of 15 June 763 BCE,
visible as a partial obscuration in Palestine (mag = 0.902),
would match the lifetime of Amos better \cite{stephenson:1975}.
Another passage is found in Joel 2:31, but the line does not seem
to relate to a concrete sighting but rather give a general scenario
of divine power.

As hinted in the historical foreword, substantial parts of the
text were composed in the 7th century and rearranged during the
Babylonian exile between 586 and 538 BCE.
This might open up the option that the author was located in Babylon
and interwove an eclipse from there.
Our check does not deliver an appropriate totality during the whole
6th century BCE though.
The closest match would be 28 May 585 BCE (Thales' eclipse) with a
mag = 0.998 occurring shortly \emph{after} the sun has set:
the maximum would have happened below the horizon.
The possibility of involving a ``contemporary eclipse'' from Babylon
must be ruled out.

The cessation of the sun is mentioned three times in the lines
quoted above.
There is a remark between the second and the third mention.
The writer gives reference to another ancient source:
the Book of Jashar.
This awkward passage stands without a connection to the incident or
any other miracle by God, thus, it remains unclear how much of the
verse is just part of a quote and whether it refers to this battle
at all \cite{creach:2003}.
The Bible contains a second reference to that Jashar (2.\ Samuel
1:18), where he is deemed ``upright''.
However, such a book does not exist \cite{w-en}.
It is completely unknown to literature.
The interposed comment must be a later amendment, and it serves like
a justification for the slaughter performed by the Jewish army
among the defeated.

Apart from many allusions to strange happenings in the sky, this
report at Gibeon seems to represent a more authentic record of a
historic kernel.
Some elements of the story were altered and embellished to cover
criminal acts.
Such an adornment might be the falling stones in verse 11, which
could be taken for meteors or a severe hailstorm.
However, nobody will ever be injured by them.
The meteor shower must have been seen on a different day, anyway.
It supports the idea that the anonymous writer combined various
events into one great spectacle to be preserved as a ``cultural
heritage'' of a people.
The verses in Joshua 10 were never meant as a thorough reflection
of facts.

\section{Summary}

We gave a resume on the Jewish history and pointed out the fakes
entering many religious narrations.
A historical kernel is often modified and embellished with
``miracles''.
The eclipse of 30 September 1131 BCE came as a surprise during the
Battle of Gibeon in Palestine and affected the outcome.

The eclipse would serve as a fixpoint for investigating the
deceleration of the Earth's rotation.
On recalculating the Besselian Elements, the value of this parameter
can be confined to an error of $\approx$400 seconds (6,5 minutes)
centred on $\Delta T$ = 27,840 s.
The account could be the oldest account of secured events unless some
doubts restrain us from so declaring:
in the 6th century BCE it was rewritten to admonish the
contemporaries to stick to older traditions.
The entire story is not supported by independent sources, and
totality is not compelling for the confusion during the battle.
After the experience of 1131 BCE no totality was recorded in
Palestine for more than 700 years.


\section*{Acknowledgements}

This article is a revised chapter of the author's Habilitation
submitted to the University of Heidelberg in February 2020.
The author expresses his deep thankfulness to his family members
and friends for their support against the obstacles caused to his
life for many, many years.


\vspace{\baselineskip}

\bibliography{v182-biblio}


\end{document}